\documentclass[preprint2]{aastex}

\newcommand{\fourvec}[1]{\mbox{\boldmath $\mathsf{#1}$}}

\shorttitle{Testing no-hair theorems}
\shortauthors{Will}

\usepackage{epsfig}

\begin{document}


\title{Testing the general relativistic ``no-hair'' theorems using the 
galactic center black hole SgrA$^*$}



\author{Clifford M. Will}
\affil{McDonnell Center for
the Space Sciences, Department of Physics, \\
Washington University, St.
Louis, Missouri 63130; cmw@wuphys.wustl.edu}


\begin{abstract}
If a class of stars orbits the central black hole in
our galaxy in short period ($\sim 0.1$ year), high eccentricity ($\sim 0.9$) 
orbits, 
they will experience
precessions of their orbital planes induced by both relativistic
frame-dragging and
the quadrupolar gravity of the hole, at levels that could be as large
as 10 $\mu$arcseconds per year, if the black hole is rotating faster
than 1/2 of its maximum rotation rate.  Astrometric observations of the
orbits of at least two such stars can in principle lead to a
determination of the angular momentum vector ${\bf J}$ of the black hole
and its quadrupole moment $Q_2$.  This could lead to 
a test of the general relativistic 
no-hair theorems, which demand that 
$Q_2 = -
J^2/M$.   Future high-precision adaptive infrared optics instruments
may make such a fundamental test of the black-hole paradigm possible. 

\end{abstract}


\section{Introduction}\label{secI}

The discovery, using infrared telescopes, 
of stars orbiting within an arcsecond of
the central object SgrA$^*$ in our galaxy, together with accurate
determinations of their orbits,
has provided strong evidence for the existence there of a massive black hole
(MBH) of around $3.6 \times 10^6 \,M_\odot$ 
(see \cite{alexander05} for a review).
In addition to opening a window on the innermost region of
the galactic center, the discovery of these stars has made it possible
to contemplate using orbital dynamics to probe the curved spacetime
of a rotating black hole.

The orbital periods of these stars
are on the scale of tens of years, and thus most relativistic effects,
such as the advance of the pericenter,
are too small to be observed at present (see, however \cite{zucker}).  
Nevertheless,
there seems to be every expectation that, with improved observing
capabilities,  a population of stars significantly closer to the hole
will eventually be discovered, making orbital
relativistic effects detectable
\citep{jaroszynski,fragile,rubilar,weinberg,kraniotis}.  Furthermore, plans are being developed
to achieve infrared astrometry on such objects
at the level of 10 $\mu$arcseconds \citep{eisenhauer}.  High-precision
Doppler measurements may also be possible \citep{zucker}.

This makes it possible to consider doing more than merely detect
relativistic effects, but rather to provide the first test of the
black hole ``no-hair'' or uniqueness theorems of general relativity.   
According to
those theorems, an electrically neutral black hole 
is completely characterized by
its mass $M$ and angular momentum $J$.  As a
consequence, all the multipole moments of its external spacetime are
functions of $M$ and $J$, specifically, the quadrupole moment $Q_2 = -J^2/M$ 
(in units where $G=c=1$).  

If the black hole were non-rotating ($J=0$), then its exterior metric
would be that of Schwarzschild, and the most important relativistic effect     
would be the advance of the pericenter.  
If it is rotating, then two new phenomena occur, the dragging of
inertial frames and the effects of the hole's quadrupole moment, 
leading not only to an additional pericenter
precession, but also to a precession of the
orbital plane of the star.  These precessions are smaller than the
Schwarzschild effect in magnitude because they depend on the
dimensionless angular momentum parameter $\chi \equiv J/M^2$, 
which is always less
than one, and because they fall off faster with distance from the
black hole.
However, accumulating evidence 
suggests that MBH should be rather rapidly 
rotating, with $\chi$ larger than 0.5 and possibly as large as 0.9, 
so these effects
could be significant.

\begin{figure}[t]
\begin{center}
\epsfig{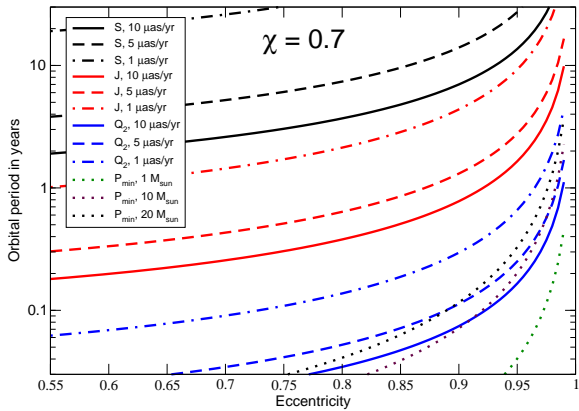}
\caption{Orbital periods vs. eccentricity required to give measurable
relativistic
precession rates.  Dotted curves show minimum
periods vs. $e$ that avoid tidal disruption, for various stellar
masses.
\label{fig1}}
\end{center}
\end{figure}

The purpose of this paper is to point out that, if a class of
stars were to be found with orbital periods of fractions of a year, and
with sufficiently 
large orbital eccentricities, then the quadrupole-induced precessions
could be as large as 10 $\mu$as per year.   Figure \ref{fig1} illustrates
this: assuming a black hole with $\chi = 0.7$, it shows the orbital
period required as a function of eccentricity, for the rates of precessions
due to
Schwarzschild (S), frame-dragging ($J$) and quadrupole ($Q_2$) terms to be as
large as 10, 5, and 1 $\mu$as per year.

\begin{figure}[t]
\begin{center}
\epsfig{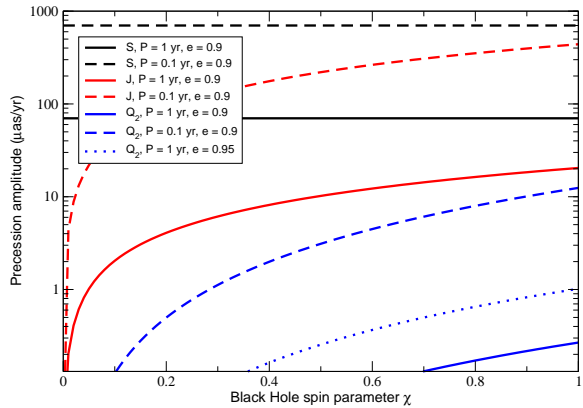}
\caption{Relativistic precession amplitudes vs. black hole spin parameter
$\chi$.
\label{fig2}}
\end{center}
\end{figure}

Figure \ref{fig2} shows the effect of black hole spin on the amplitudes of
the relativistic effects.  For orbits with eccentricity 0.9 and periods
of one year and 0.1 years, the amplitudes of the three effects are
plotted in $\mu$as per year.    

The precession of the orbital plane is the most important effect here,
because it depends only on $J$ and $Q_2$; the Schwarzschild
part of the
metric affects only the pericenter advance.  
The orbital plane is determined by its inclination angle $i$ relative
to the plane of the sky and by the angle of nodes $\Omega$ 
between a reference direction and the
intersection of the 
two planes.  Standard astrometric and Doppler observations
can determine $\Omega$,
$i$, the pericenter angle $\omega$, the semimajor axis $a$ and the
orbital eccentricity $e$, and, given sufficient observation time, the
secular rates of change $d\Omega/dt$, $di/dt$, and $d\omega/dt$.  

However, in order to test the no-hair theorems, one must determine
five parameters: the mass of the black hole, 
the magnitude and two angles of its
spin, and the value of the
quadrupole moment.  
The ``Kepler-measured'' mass is determined from the orbital periods of 
stars, but may require
data from a number of stars to fix it separately from any extended
distribution of mass.  
Then, to measure $\bf J$ and $Q_2$, it is necessary and 
sufficient to measure
$d\Omega/dt$ and  $di/dt$ for two stars in non-degenerate orbits.  
A test of the no-hairness of
the central object in our galaxy would be compelling evidence that it
is truly a black hole of general relativity.

\section{Orbit perturbations in the field of a rotating black hole}\label{secII}

For the purpose of this rough analysis, it suffices to work in the
post-Newtonian limit.  The
equation of motion of a body of negligible mass in the field of a body
with mass $M$, angular momentum $\bf J $ and quadrupole moment $Q_2$ is
given by
\begin{eqnarray}
{\bf a} &=& -\frac{M{\bf x}}{r^3} + \frac{M{\bf x}}{r^3} \left (
4 \frac{M}{r} - v^2 \right ) +4\frac{M{\dot r}}{r^2} {\bf v}
\nonumber \\
&&
- \frac{2J}{r^3} [ 2{\bf v}\times {\bf {\hat J}}
-3 {\dot r} {\bf n}\times {\bf {\hat J}} - 3{\bf n}({\bf h} \cdot
{\bf {\hat J}})/r ] 
\nonumber \\
&&
+ \frac{3}{2} \frac{Q_2}{r^4} [5{\bf n}({\bf n} \cdot {\bf {\hat J}})^2
- 2 ({\bf n} \cdot {\bf {\hat J}}){\bf {\hat J}} - {\bf n} ] \,,
\label{eom}
\end{eqnarray}
~where $\bf x$ and $\bf v$ are the position and velocity of the
body, ${\bf n} = {\bf x}/r$, ${\dot r} = {\bf n}\cdot {\bf v}$,  
${\bf h}= {\bf x} \times {\bf v}$, 
and
${\bf {\hat J}} = {\bf J}/|J|$ (see, eg. \cite{tegp}).   
The first line of Eq.
(\ref{eom})
corresponds to the Schwarzschild part of the metric (at post-Newtonian
order), the second line is the frame-dragging effect, and the third
line is the effect of the quadrupole moment (formally a
Newtonian-order effect).  
For an axisymmetric black hole, the symmetry axis of the hole's
quadrupole moment coincides with its rotation axis,
given by the unit vector ${\bf {\hat J}}$.  The
magnitude of the quadrupole moment will be left free.

Using standard orbital perturbation theory, we find that the precessions
per orbit of  the orientation variables are given by
\begin{mathletters}
\begin{eqnarray}
\Delta {\varpi} &=& A_S - 2 A_J \cos \alpha
\nonumber
\\
&& - \frac{1}{2} A_{Q_2}
(1-3\cos ^2 \alpha ) \,,
\label{domega}
\\
\sin i \Delta \Omega &=& \sin \alpha \sin \beta (A_J - A_{Q_2} \cos \alpha) \,,
\label{dOmega}
\\
\Delta i  &=& \sin \alpha \cos \beta (A_J - A_{Q_2} \cos \alpha) \,,
\label{di}
\end{eqnarray}
\label{delements}
\end{mathletters}
~where
\begin{mathletters}
\begin{eqnarray}
A_S &=& 6\pi M/p \,,
\\
A_J &=& 4\pi J/(mp^3)^{1/2} \,,
\\
A_{Q_2} &=& 3\pi Q_2/mp^2 \,,
\label{amplitudes}
\end{eqnarray}
\end{mathletters}
~where $\Delta {\varpi} = \Delta \omega + \cos i \,\Delta \Omega$ is
the precession of pericenter relative to the fixed reference direction,
and $p=a(1-e^2)$ is the semilatus rectum.  
The quantities $\alpha$ and $\beta$ are the 
polar angles
of the black hole's angular momentum vector with respect the star's
orbital
plane defined by
the line of nodes ${\bf e}_p$, and the vector in the
orbital plane ${\bf e}_q$ orthogonal to ${\bf e}_p$ and ${\bf h}$. 

The structure of Eqs. (\ref{dOmega}) and (\ref{di}) can be
understood as follows: Eq. (\ref{eom}) implies that the orbital
angular momentum ${\bf h}$ precesses according to $d{\bf h}/dt =
\fourvec{\omega} \times {\bf h}$, where the orbit-averaged
$\fourvec{\omega}$ is given by
$\fourvec{\omega} = {\bf {\hat J}} (A_J - A_{Q_2} \cos \alpha)$;
the orbit element variations are given by 
$di/dt = \fourvec{\omega} \cdot {\bf e}_p$ and
$\sin i d\Omega/dt =\fourvec{\omega} \cdot {\bf e}_q$.
As a consequence, we have the purely geometric relationship,
\begin{equation}
\frac{\sin i d\Omega/dt }{di/dt} = \tan \beta
\,.
\label{tanbeta}
\end{equation}

To get an idea of the
astrometric size of these precessions, we define an angular precession
rate amplitude ${\dot \Theta}_i = (a/D) A_i/P$, where $D$ is the
distance to the galactic center and $P=2\pi (a^3/M)^{1/2}$ is the
orbital period.  Using $M=3.6 \times 10^6 \, M_\odot$, $D = 8 \, {\rm
kpc}$, we obtain the rates, in
microarcseconds per year
\begin{eqnarray}
{\dot \Theta}_S &\approx& 13.3 \,  P^{-1} (1-e^2)^{-1}
\,,  
\\
{\dot \Theta}_J &\approx& 0.847 \,  \chi P^{-4/3}
(1-e^2)^{-3/2}  
\,, 
\\
{\dot \Theta}_{Q_2} &\approx& 9.68 \times 10^{-3}\, 
\chi^2
P^{-5/3}
(1-e^2)^{-2}  
\,, 
\end{eqnarray}
~where we have assumed
$|Q_2|= M^3 \chi^2$. 
The observable precessions will be
reduced somewhat from these raw rates because the orbit must be
projected onto the plane of the sky.  For example, the contributions
to $\Delta i$ and $\sin i \Delta \Omega$ are reduced by a factor $\sin
i$; for an orbit in the plane of the sky, the plane precessions are
unmeasurable.
 
For the quadrupole precessions to be observable, it is clear that the
black hole must have a decent angular momentum ($\chi > 0.5$) and that
the star must be in a short period high-eccentricity orbit.  Figures
\ref{fig1} and \ref{fig2} show the quantitive requirements,
based on these rate amplitudes.

\section{Testing the no-hair theorems}\label{secIII}

Although the pericenter advance is the largest relativistic orbital
effect, it is {\it not} the most suitable effect for testing the
no-hair theorems.  The frame-dragging and quadrupole effects are small
corrections of the leading Schwarzschild pericenter precession, and thus one
would need to know $M$, $a$ and $e$ to sufficient accuracy to be able
to subtract
that dominant term to reveal the smaller effects of interest.
Furthermore, the pericenter advance is affected by a number of
complicating phenomena. (i) For such relativistic orbits,
Schwarzschild
contributions to the pericenter precession at the {\em second}
post-Newtonian order may be needed. (ii) Any distribution of mass
(such as dark matter or gas) within the orbit, even if it is
spherically symmetric, will generally contribute to the pericenter advance.
(iii) Tidal distortions of the stars are likely to occur near the
pericenters of the highly eccentric orbits, leading to additional
contributions to the pericenter advance of the form 
$30\pi (M/m)(R/a)^5 k_2
(1+3e^2/2+e^4/8)/(1-e^2)^5$, where $m$, $R$ and $k_2$ are the mass,
radius and ``apsidal constant'', or Love number of the star,
respectively.  Tidal contributions could be significant for sufficiently
close and eccentric orbits.  

Of course, if a star gets too close to
the black hole, it could be tidally disrupted.  This possibility sets a
lower bound on the orbital period $P_{\rm min} \sim 2\sqrt{3}\pi
(R^3/m)^{1/2}(1-e)^{-3/2}$, set by requiring that the pericenter
distance exceed the Roche radius of the star.  This is
illustrated by the dotted curves in Fig. \ref{fig1}.

By contrast, the precessions of the node and inclination are
relatively immune from such effects.  Any spherically symmetric
distribution of mass has no effect on these orbit elements.  As long
as any tidal distortion of the star is  quasi-equilibrium 
with negligible tidal lag, the resulting perturbing forces are purely
radial, and thus have no effect on
the node or inclination.  

From the measured orbit elements 
and their drifts for a given star, 
Eq. (\ref{tanbeta}) gives the angle $\beta$, 
independently of any assumption about the no-hair theorems.  This
measurement then fixes the spin axis of the black hole to lie on a plane
perpendicular to the star's orbital plane that makes an angle $\beta$
relative to the line of nodes.
The equivalent determination for another stellar orbit fixes another
plane; as long as the two planes are not degenerate, their
intersection determines the direction of the spin axis, modulo a
reflection through the origin.  

This information is then sufficient to determine the angles 
$\alpha$ and $\beta$
for each star.  
Then, from the magnitude
\begin{equation}
\left ( [\sin i \frac{d\Omega}{dt}]^2 + [\frac{di}{dt}]^2 \right )^{1/2}
= \sin \alpha (A_J - A_{Q_2} \cos \alpha) \,,
\end{equation}
determined for each star, together with the orbit elements, 
one can solve for $J$ and
$Q_2$.

In practice, of course, the analysis of the astrometric data will be
carried out in a more sophisticated, if less transparent manner.  
Using data from all detected stars,  one 
carries out a multi-parameter least-squares fit, 
standard in solar-system celestial mechanics, to determine their orbit
elements.  Their motions would be based on Eq. (\ref{eom}) but with
$M$, ${\bf J}$ and $Q_2$ treated as parameters to be fit along with the
orbit elements of each star.
If necessary, the
model can be extended to include effects of an additional matter
distribution, tidal effects, and so on.

\section{Concluding remarks}
\label{secV}

We have shown that a class of stars orbiting a rotating central black hole in
our galaxy in short period, high eccentricity orbits, will experience
precessions of their orbital planes induced by both frame dragging and
the quadrupolar gravity of the hole, at levels that could be as large
as 10 $\mu$arcseconds per year.  Observations of the
orbits of at least two such stars can in principle lead to a
determination of the angular momentum vector ${\bf J}$ and quadrupole
moment $Q_2$ of the black hole, and could provide a test of the 
no-hair theorems of general relativity.

Alternative possibilities for no-hair tests involve timing measurements
of pulsars orbiting black-hole companions \citep{kopeikin}, 
gravitational-wave measurements of compact objects spiralling into massive
black holes \citep{ryan,glampedakis,hughes}, or detection of quasi-normal
``ringdown'' gravitational
radiation of perturbed black holes \citep{dreyer,bcw}. 

Detecting such stars so close to the black hole, and carrying out
infrared astrometry to $10 \, \mu$arcsecond accuracy will be a
challenge.  However, if this challenge can be met with future improved
adaptive optics systems currently under study, such as
GRAVITY \citep{eisenhauer}, it could lead to a powerful test of the black-hole
paradigm.

In future work, we plan to study in detail such
complicating effects as second-post-Newtonian (2PN) corrections to the
Schwarzschild part of the pericenter advance, tidal effects, effects
of unseen mass distributions within the observed stellar orbits,
and light
deflection and Shapiro time delay effects \citep{rubilar,weinberg}. 
For example, a torus of matter of mass $m$ orbiting the black hole at
a distance $R$ will induce fractional changes in the apparent angular
momentum and quadrupole moment of order $\delta J/J \sim
(m/M)(R/M)^{1/2}(1/\chi)$ and $\delta Q/Q \sim (m/M)(R/M)^2
(1/\chi)^2$, so only  a very massive and/or very distant torus
will be relevant.
We also plan to carry out covariance analyses to obtain more realistic
estimates of the accuracies that might be obtained for the no-hair
test
for given raw astrometric accuracies, and for a range of
observing schedules.

\acknowledgments 
This work was supported in part by the National Science Foundation
under grant No. PHY 06-52448.  
We  are grateful to the  Groupe Gravitation et Cosmologie
(GR$\varepsilon$CO),
Institut d'Astrophysique de Paris, Universit\'e Pierre et Marie Curie
for their hospitality
during the initial stages of this work.  Peter Ronhovde made useful
contributions at an early phase of this study.  We thank Andrea Ghez
and Robert Reasenberg for helpful comments.


\begin{thebibliography}{}

\bibitem[Alexander(2005)]{alexander05}
Alexander, T. (2005), Phys. Rep., 419, 65

\bibitem[Berti, et al.(2006)]{bcw}
Berti, E., Cardoso, V. \& Will, C. M. (2006), Phys. Rev. D, 73, 064030

\bibitem[Dreyer, et al.(2004)]{dreyer}
Dreyer, O., Kelly, B., Krishnan, B., Finn, L. S., Garrison, D. \&
Lopez-Aleman, R. (2004), Class. Quantum. Grav. 21, 787

\bibitem[Eisenhauer et al.(2008)]{eisenhauer}
Eisenhauer, F., Perrin, G., Rabien, S., Eckart, A., L\'ena, P.,
Genzel, R., Abuter, R., Paumard, T., \& Brandner, W. (2008), in 
The Power of Optical/IR
Interferometry: Recent Scientific Results and 2nd Generation 
Instrumentation, Eds. A. Richichi, F.
Delplancke, F. Paresce, \& A. Chelli, Berlin: Springer-Verlag 

\bibitem[Fragile \& Mathews(2000)]{fragile}
Fragile, P. C. \& Mathews, G. J. (2000), \apj, 542, 328

\bibitem[Glampedakis \& Babak(2006)]{glampedakis}
Glampedakis, K. \& Babak, S. (2006), Class. Quantum Grav., 23, 4167

\bibitem[Hughes(2006)]{hughes}
Hughes, S. A. (2006), in Laser Interferometer Space Antenna, 6th
International LISA Symposium, Eds. S. M. Merkowitz, J. C. Livas, New
York: AIP Conference Proceedings, Vol. 873, p. 233
 
\bibitem[Jaroszy\'nski(1998)]{jaroszynski}
Jaroszy\'nski, M. (1998), Acta Astronomica, 48, 653

\bibitem[Kraniotis(2007)]{kraniotis}
Kraniotis, G. V. (2007), Class. Quantum Grav., 24, 1775

\bibitem[Rubilar \& Eckart(2001)]{rubilar}
Rubilar, G. F., \& Eckart, A. (2001), A \& A, 374, 95

\bibitem[Ryan(1997)]{ryan}
Ryan, F. D. (1997), Phys. Rev. D, 56, 1845
\bibitem[Weinberg et al.(2005)]{weinberg}
Weinberg, N. N., Milosavljevi\'c, M., \& Ghez, A. M. (2005), \apj,
622, 878

\bibitem[Wex \& Kopeikin(1999)]{kopeikin}
Wex, N. \& Kopeikin, S. M. (1999), \apj, 514, 388

\bibitem[Will(1993)]{tegp}
Will, C. M. (1993), Theory and Experiment in Gravitational Physics,
2nd Ed., Cambridge: Cambridge University Press

\bibitem[Zucker et al.(2006)]{zucker}
Zucker, S., Alexander, T., Gillessen, S., Eisenhauer, F., \& 
Genzel, R. (2006), \apjl, 639, L21

\end{thebibliography}
\end{document}